\documentclass[]{aa} 
\pdfoutput=1

\usepackage{graphicx}
\usepackage[colorlinks=true,pdfborder={0 0 0},linkcolor=blue,citecolor=blue]{hyperref}

\usepackage[varg]{txfonts}
\begin{document}

   \title{Magnetic field inference from the spectral region around the Mg\,I $b_2$ line using the weak-field approximation}

   \author{D. Vukadinovi\'c
          \inst{1,2}
          ,
          I. Mili\'c
          \inst{3,4,5,6}
          and
          O. Atanackovi\'c
          \inst{1}
          }

   \institute{Department of Astronomy, Faculty of Mathematics, University of Belgrade, Studentski trg 16, Belgrade
         \and
             Max Planck Institute for Solar system research, Justus-von-Liebig Weg 3, Goettingen \\
             \email{vukadinovic@mps.mpg.de}
         \and
            Department of Physics, University of Colorado, Boulder CO 80309, USA
         \and
            Laboratory for Atmospheric and Space Physics, University of Colorado, Boulder CO 80303, USA
         \and
            National Solar Observatory, Boulder CO 80303, USA
        \and
            Astronomical observatory Belgrade, Volgina 7, 11060 Belgrade, Serbia
             }

   \date{Received month dd, yyyy; accepted month dd, yyyy}
   
\authorrunning{Vukadinovi\'{c} et al.}
\titlerunning{Magnetic field inference using the spectral region around Mg\,I $b_2$ line using the weak-field approximation}

\abstract
{The understanding of the magnetic field structure in the solar atmosphere is important for assessing both dynamics and energy balance of the solar atmosphere. Our knowledge about these magnetic fields comes predominantly from the interpretation of spectropolarimetric observations. Simpler, approximate approaches like the weak-field approximation (WFA) deserve special attention as the key methods for the interpretation of large, high-resolution datasets.}
{We investigate the applicability of the WFA for retrieving the depth-dependent line-of-sight (LOS) magnetic field from the spectral region containing Mg\,I $b_2$ spectral line and two photospheric Ti\,I and Fe\,I lines in its wings.}
{We constructed and used a 12-level model for Mg\,I atom that realistically reproduces the $b_2$ line profile of the mean quiet Sun. We tested the applicability of the WFA to the spectra computed from the FAL\,C atmospheric model with ad hoc added different magnetic and velocity fields. Then, we extended the analysis to the spectra computed from two 3D magneto-hydrodynamic (MHD) MURaM simulations of the solar atmosphere. The first MHD cube is used to estimate the Stokes $V$ formation heights of each spectral line. These heights correspond to optical depths at which the standard deviation of the difference between the WFA-inferred magnetic field and the magnetic field in the MHD cube is minimal. The estimated formation heights are verified using the second MHD cube.}
{The LOS magnetic field retrieved by the WFA is reliable for the magnetic field strength up to $1.4$\,kG even when moderate velocity gradients are present. The exception is the Fe\,I line, for which we found a strong discrepancy in the WFA-inferred magnetic fields because of the line blend. We estimated the Stokes $V$ formation heights of each spectral line to be:  $\log\tau_\mathrm{Fe}=-2.6$, $\log\tau_\mathrm{Mg}=-3.3$, and $\log\tau_\mathrm{Ti}=-1.8$. We were able to estimate the LOS magnetic field from the MURaM cube at these heights with the uncertainty of $150$\,G for the Fe\,I and Ti\,I lines and only $40$\,G for the Mg\,I $b_2$ line.}
{Using the WFA we can quickly get a reliable estimate of the structure of the LOS magnetic field in the observed region, which is a significant advantage in comparison with time consuming classical spectropolarimetric inversions. The Mg\,I $b_2$ line profile calculated from the quiet Sun MURaM simulation agrees very well with the observed mean spectrum of the quiet Sun.}


   \maketitle
%
\section{Introduction}

Estimating the magnetic fields in the atmosphere of the Sun is essential for studying solar activity, modeling space weather, and understanding the coupling between different layers of the solar atmosphere. Especially in the chromosphere, gas pressure is low enough that the magnetic field makes a significant contribution to the energy balance. \cite{MartinezSykora19} showed that the net magnetic flux from a simulated quiet photosphere is not sufficient to maintain the chromospheric magnetic field and that different types of magnetic structures can be formed in situ as a result of chromospheric dynamics. To fully understand the structure and the energy flow in the solar atmosphere, it is important to determine the structure of the magnetic field, i.e. to infer its spatial variation, preferably in all three dimensions.

Solar magnetic fields are most often inferred from spectropolarimetric observations. The depth distribution of the magnetic field (and other physical quantities) in the atmosphere of the Sun can be derived by analyzing the spectral lines that form at various depths. High-angular-resolution observations allow us to study the variation of physical parameters over the solar surface. Therefore, the observations with high spectral and spatial resolution allow us to perform three-dimensional mapping of the solar atmosphere. State-of-the-art solar telescopes, such as Swedish Solar Telescope \citep{sst}, SUNRISE \citep{sunrise}, and DKIST \citep{dkist} are designed with this kind of studies in mind.

Interpreting spectropolarimetric observations requires the modeling of spectra, i.e. relating the atmospheric conditions to the emergent polarized spectra. The most common approach is spectropolarimetric inversion \citep{inversion} that uses different minimization strategies to find the best fit of the atmospheric model to the observed Stokes spectrum. The parameters of atmospheric model are modified in each iteration, making the process computationally demanding, and especially so for the modeling of chromospheric spectral lines. At low chromospheric densities the treatment of a non-local and non-linear coupling between the plasma and the radiation field (non-local thermodynamic equilibrium regime, NLTE) increases the computational cost by several orders of magnitude. Numerical costs become even higher when high-angular-resolution spectropolarimetric observations are analyzed. Therefore, NLTE spectral line inversion codes \citep[e.g.][]{snapi, stic} might not be suitable for the analysis of enormous data sets provided by the next-generation instruments.

A simple and efficient alternative to NLTE spectral line inversion codes is to employ the weak-field approximation (WFA) to estimate the magnetic field from spectropolarimetric observations. This approximation holds when the Zeeman splitting of a spectral line is much smaller than its Doppler width. By perturbing the polarized radiative transfer equation, one can derive expressions that relate the circular and linear polarization signals to the first and the second derivatives of the intensity with respect to wavelength through the magnetic field\citep[e.g.][]{biblija, rebeka}. The WFA is derived under the assumption of constant magnetic and velocity fields throughout the observed atmosphere. It does not depend on the line formation mechanism, so it can be applied to any spectral line of interest. Each specific spectral line yields a unique value of the line-of-sight (LOS) or the transversal magnetic field, that corresponds to some sort of mean magnetic field in the line forming region \citep{Borrero2014,inversion}.

The chromospheric magnetic field is commonly probed by using the Ca\,II $8542$ line \citep[e.g.][]{QuinteroNoda19,Pietrow20}. However, there are only a few investigations that focus on the layers between the photosphere and the chromosphere, i.e. the region of the so-called "temperature minimum". Diagnostically suitable spectral line that probes this region is the Mg\,I $b_2$ line \citep{Lites88,Quintero_mg}. Out of the three Mg\,I $b$ lines: $b_1$ at $5183.6$\,\AA, $b_2$ at $5172.7$\,\AA\ and $b_4$ at $5167.3$\,\AA, $b_2$ line is the most convenient because it has the strongest polarization signal (Land{\' e} factor $g_\mathrm{eff} = 1.75$), and does not contain blends. This line was observed by the NFI instrument on-board the HINODE solar telescope \citep{Hinode04} and will be observed by the TuMag instrument on-board the balloon-borne SUNRISE mission \citep[][]{Quintero_mg}.

In this paper we examine the diagnostic capabilities of the spectral region around the Mg\,I $b_2$ line for the magnetic field inference using the WFA. We focus on the $b_2$ line and two neighboring photospheric spectral lines, Fe\,I $5171.61$ and Ti\,I $5173.75$, where the former is a blend of two neutral iron lines. Our goal is to infer the depth dependent LOS magnetic field strength by applying the WFA to the synthetic polarized spectra of these three spectral lines. For this, we use two MURaM quiet Sun simulations \citep{muram} that reproduce observed properties of the Sun to a high degree \citep{Sanja10}. We synthesize the polarized spectra from one MURaM simulation, apply the WFA to each of the three lines, and compare the results with the original magnetic field in the simulation. This allows us to understand what depths the spectral lines are probing. We then test the inferred depths by applying the WFA to the second MURaM simulation and repeat the comparison.

The outline of the paper is as follows: In section 2 we construct a reliable and robust model of Mg\,I atom that reproduces well the mean solar spectrum of the $b_2$ line and is simple enough to enable fast synthesis under the NLTE conditions. In section 3 we discuss the validity of the weak-field approximation for the lines of interest in one-dimensional atmospheric models, following the approach of \cite{rebeka}. In section 4 we propose and test the method to infer the depth-dependent LOS magnetic field using synthetic data from MURaM simulations, and present our conclusions in section 5.

\section{Spectral synthesis}

Our first goal is to calculate spectral region around the Mg\,I $b_2$ line that matches well the mean observed spectrum of the quiet Sun. To model a spectral line in NLTE conditions, it is necessary to construct an adequate atomic model that takes into account all the relevant processes contributing to the level populations of the line transition under study. In local thermodynamic equilibrium (LTE), atomic level populations follow from the Saha-Boltzmann distribution, while in NLTE we need to solve the statistical equilibrium equations self-consistently with radiative transfer equation for the relevant transitions \citep{Mihalas}. 

\begin{figure}
   \includegraphics[width=\hsize]{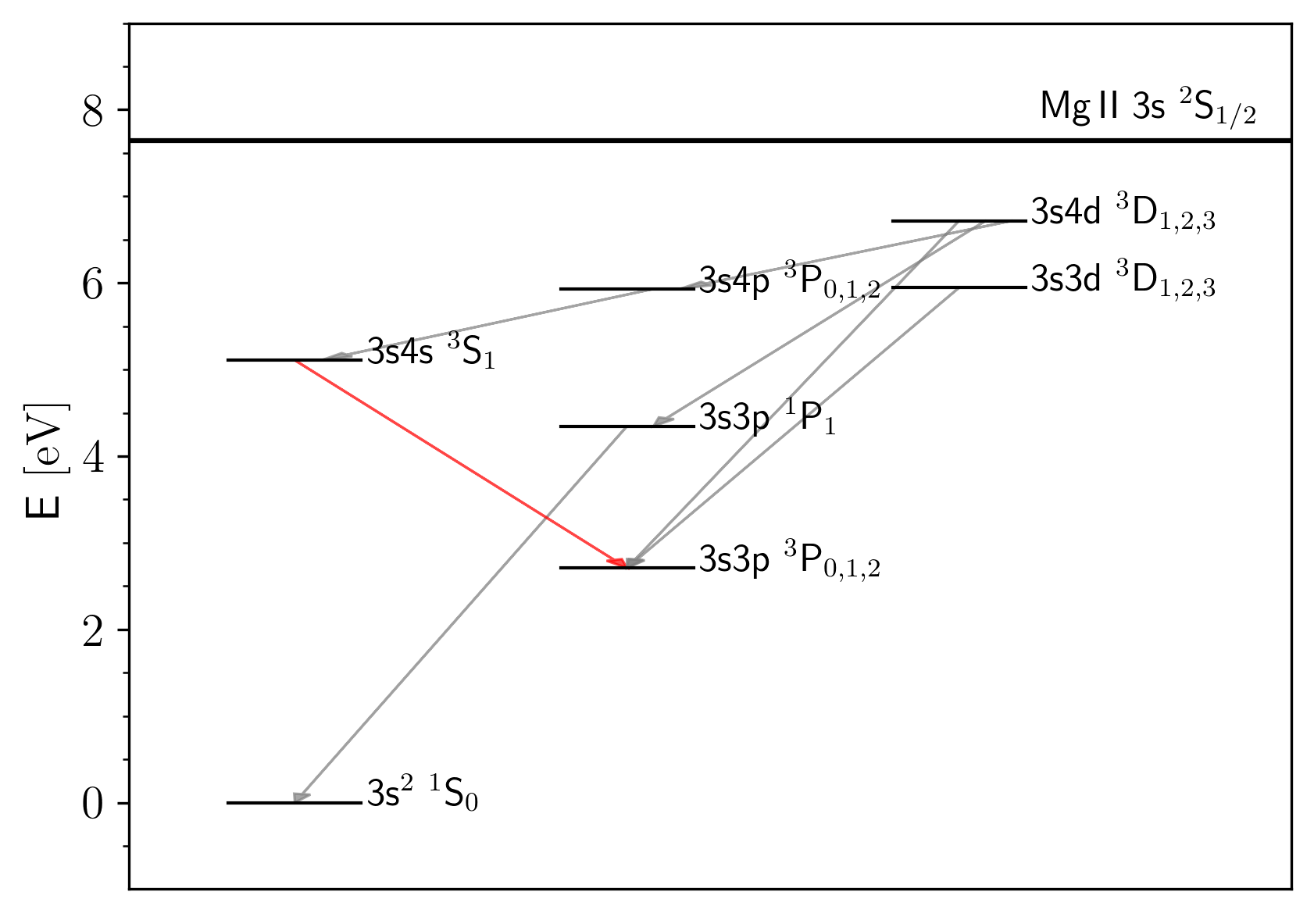}
   \caption{The schematic term diagram of the 12-level Mg\,I atomic model. Red line denotes the transition in which the Mg\,I $b$ lines are formed. Full lines represent the bound-bound transitions. From the each bound level we have a bound-free transition.}
   \label{fig:grotrian}
\end{figure}

One of the first studies of the Mg\,I $b_2$ line was done by \cite{mauas} who discussed the influence of different energy levels and collision coefficients on the line strength. They concluded that the inclusion of triplet levels in the atomic model has a significant influence to the line strength. Following their work, we constructed a 12-level atomic model (11 energy levels of Mg\,I plus the Mg\,II continuum, see Fig.\,\ref{fig:grotrian}).

Atomic data for the considered energy levels are given in Table \ref{tab:mg_atom}. Since higher atomic levels are very close in energy, we merged them into superlevels so that the energy of each superlevel is equal to the mean energy of the corresponding sublevels and its statistical weight is equal to the sum of all sublevel weights. The radiative de-excitation rate is set to the statistical weight-weighted mean of all the sublevels. Merging levels into superlevels decreases the total number of transitions, thus decreasing computing time and increasing numerical stability of the solution. \cite{Quintero_mg} used a very similar atomic model with 13 levels, out of which 7 levels were treated as superlevels. Compared to more complicated atomic models \citep[e.g.][]{osorio} our model allows for faster synthesis, while producing line shapes that agree well with the observations.

\begin{table}[hbt!]
    \caption{The energy levels of Mg\,I atomic model. Data given in table are: level number, level term, total angular momentum ($J$), energy of the level ($E$), statistical weight ($g$), and Land{\' e} factor ($g_L$). All data are from the NIST database \citep{NIST} while the Land{\' e} factors are from \cite{Kurucz}.}
    \begin{tabular}{c c c c c c c}
    \hline \hline
        \# & Term & J & $E$ [eV] & $g$ & $g_L$ \\ \hline
        1 & 3s$^2$ $^1$S & 0 & 0 & 1 & 0 \\
        2 & 3s3p $^3$P$^\circ$ & 0 & 2.7091 & 1 & 0  \\
        3 & & 1 & 2.7115 & 3 & 1.6875 \\
        4 & & 2 & 2.7166 & 5 & 1.5625 \\
        5 & 3s3p $^1$P$^\circ$ & 1 & 4.3458 & 3 & 1.1875 \\
        6 & 3s4s $^3$S & 1 & 5.1078 & 3 & 1.6874 \\
        7 & 3s4p $^3$P$^\circ$ & 0 & 5.9315 & 1 & 0 \\
        8 & & 1 & 5.9319 & 3 & 1.6875 \\
        9 & & 2 & 5.9327 & 5 & 1.5625 \\
        10 & 3s3d $^3$D & 0 & 5.9459 & 15 & 0 \\
        11 & 3s4d $^3$D & 0 & 6.7189 & 15 & 0 \\
        12 & 3s $^2$S & 1/2 & 7.6457 & 2 & 0 \\
    \hline
    \end{tabular}
    \label{tab:mg_atom}
\end{table}

\begin{table}[]
    \caption{The relevant transitions between the levels given in Table \ref{tab:mg_atom}.}
    \begin{tabular}{c c c c}
    \hline \hline
        Transition & $\lambda$ [nm] & A$_{ji}$ [s$^{-1}$] & $\log (gf)$ \\ \hline
        1-5 & 285.2964 & 4.91$\cdot$10$^8$ & 0.255 \\
        2-6 & 516.7322 & 1.13$\cdot$10$^7$ & -0.87 \\
        3-6 & 517.2684 & 3.37$\cdot$10$^7$ & -0.393 \\
        4-6 & 518.3604 & 5.61$\cdot$10$^7$ & -0.167 \\
        6-7 & 1504.0246 & 1.34$\cdot$10$^7$ & 0.135 \\
        6-8 & 1504.7714 & 1.34$\cdot$10$^7$ & -0.341 \\

        2-10 & 383.0443 & 5.993$\cdot$10$^6$ & -1.403 \\
        3-10 & 383.3298 & 3.808$\cdot$10$^7$ & -0.599 \\
        4-10 & 383.9352 & 6.859$\cdot$10$^7$ & -0.342 \\

        2-11 & 309.6171 & 2.06$\cdot$10$^6$ & -2.529 \\
        3-11 & 309.8093 & 1.164$\cdot$10$^7$ & -1.299 \\
        4-11 & 310.2000 & 2.113$\cdot$10$^7$ & -0.817 \\
        5-11 & 522.2925 & 1.51$\cdot$10$^4$ & -3.501 \\
        7-11 & 1576.6534 & 3.67$\cdot$10$^5$ & -1.864 \\
        8-11 & 1577.4719 & 2.306$\cdot$10$^6$ & -0.588 \\
        9-11 & 1579.1505 & 4.204$\cdot$10$^6$ & -0.105 \\
    \hline
    \end{tabular}
    \label{tab:transitions}
\end{table}

Since magnesium has a low ionization energy, at photospheric and chromospheric temperatures neutral magnesium is a minority species. This makes the modeling of photoionization of Mg\,I a critical point in reproducing the observed line profiles. To properly model the ionization balance, we have to include a large number of energy levels that are close to the continuum, and to treat the sources of opacity in UV in detail, in order to accurately describe the ionizing radiation field. Both are at odds with our need for a simple atomic model. Instead, we use the opacity fudge correction that tweaks the continuum \citep{of_corr} and the line \citep{busa} opacities in the UV part of the spectrum. This allows us to obtain a satisfactory UV ionizing radiation field without modeling the UV spectrum in great detail. To calculate photoionization rates, we use cross-sections from the TOP database \citep{TOPbase_a, TOPbase_b}.

To synthesize spectra from a given atmosphere model we used the SNAPI code \citep{snapi}, which can also perform inversions using analytically derived response functions \citep{rf_analytical}. SNAPI solves the equations of radiative transfer and statistical equilibrium in 1D plane-parallel geometry using multilevel accelerated lambda iteration \citep[MALI,][]{RH91}. In the code, the complete frequency redistribution (CRD) is assumed, which is a good approximation for the Mg\,I $b_2$ line \citep[see][]{Quintero_mg}. We treated the $b_2$ line in NLTE and the Ti\,I and Fe\,I lines assuming LTE. 

The collisional cross-sections are taken from \cite{osorio}. The line damping originating from collisions with electrons and hydrogen atoms was taken into account using the approach from \citet{Anstee95,Barklem97,Barklem98}. When solving the statistical equilibrium equations, we neglect the influence of the magnetic field on level populations (field-free regime). Spectral line polarization is calculated accounting for the Zeeman effect and neglecting the scattering polarization and the Hanle effect.

\begin{figure}
   \includegraphics[width=\hsize]{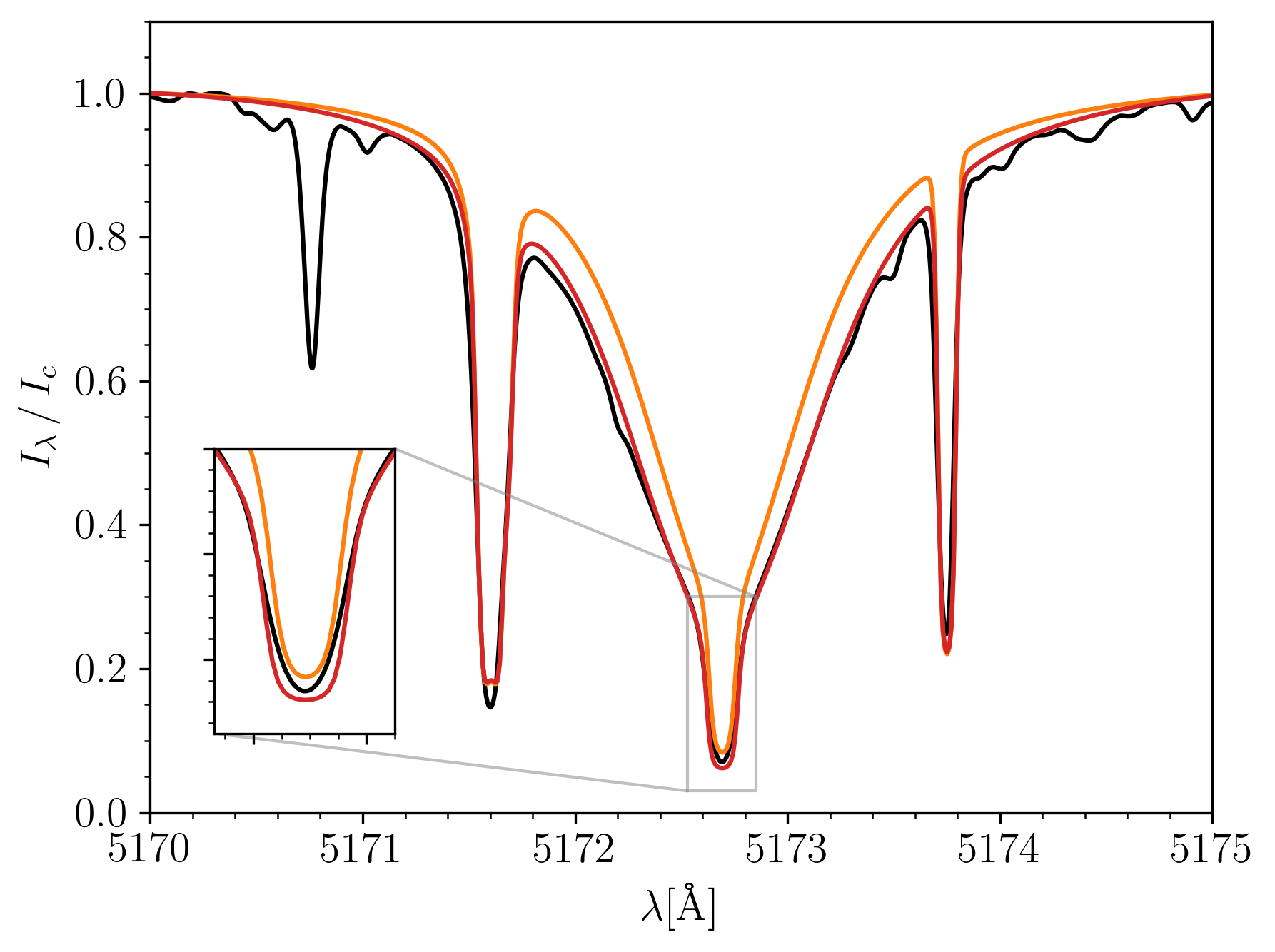}
      \caption{The comparison of the synthesized spectral region around Mg\,I $b_2$ line with (red line) and without (orange line) the opacity fudge correction, calculated from the  FAL\,C model with the mean observed solar spectrum (black line) taken from \cite{bass2000}.}
    \label{fig:of_corr}
\end{figure}

To test the constructed atomic model and the opacity fudge correction, we compared the spectrum calculated from a semi-empirical solar atmosphere model, FAL\,C \citep{fal90}, with the mean observed solar spectrum taken from \cite{bass2000} (Fig.\,\ref{fig:of_corr}). This comparison shows that we can sufficiently well reproduce the mean observed spectrum of the quiet Sun. Opacity fudge is necessary because it decreases the strength of the UV radiation, thus increasing the number of neutral magnesium atoms and resulting in a stronger Mg\,I $b_2$ line. There are small differences in the line core which can be explained by insufficiently complex atomic model. Adding a few higher levels produces a slightly better match, but at the cost of a significant increase in computing time. 

We note a pitfall of using opacity fudge correction: it is tuned to reproduce the line shapes calculated using the FAL\,C atmosphere model \citep[][]{of_corr}. Therefore, it is not adequate for other atmospheric models. We use the same fudge correction for the calculations throughout the paper, but note that a more rigorous step would be to construct a grid of opacity-fudge tables for various types of referent atmospheric models.

\section{Weak-field approximation validity in 1D atmospheres}

\label{sec:wf}

Following the study by \cite{rebeka}, we first examine the validity of the WFA for inferring the LOS magnetic field strength from the spectral region around the Mg\,I $b_2$ line. We use all three lines (Mg\,I, Fe\,I and Ti\,I) that form in distinct atmospheric layers to test how well we can retrieve the magnetic field introduced ad hoc in the semi-empirical FAL\,C atmospheric model. This will help us to better understand the inference of the LOS magnetic field from the MURaM cubes (sec. \ref{sec:wfa_3d}).

In the weak-field regime, when the Zeeman splitting of a spectral line is much smaller than its Doppler width, a perturbative scheme to the polarized radiative transfer equation can be applied, and assuming a constant magnetic field the following relationship between the circular polarization signal, V, and the strength of the LOS magnetic field, B$_\mathrm{LOS}$, can be derived \citep{delToro}:

\begin{equation}
V = - 4.6686\,10^{-13} f \lambda_0^2\ g_\mathrm{eff}\ B_\mathrm{LOS}\ \frac{\mathrm{d}I}{\mathrm{d}\lambda}.
\label{WFeq}
\end{equation}
Here, $\frac{\mathrm{d}I}{\mathrm{d}\lambda}$ is the first derivative of the intensity profile with respect to wavelength, $\lambda_0$ is the central wavelength of spectral line expressed in $\AA$, $g_\mathrm{eff}$ is Land\'e factor of the line and the LOS magnetic field $B_\mathrm{LOS}$ is expressed in Gauss. Here, the magnetic filling factor $f$ is assumed to be 1.

The validity of the WFA was tested on synthetic spectra obtained with the SNAPI code. For several constant vertical magnetic fields, ranging from $500$\,G to $2\,000$\,G in steps of $100$\,G, added to  the FAL\,C model, we analyzed the resulting synthetic Stokes profiles, derived the magnetic field strength using the WFA and compared the results to the input magnetic field of the model (see Fig.\,\ref{fig:wf_validity}).

\begin{figure}
    \includegraphics[width=\hsize]{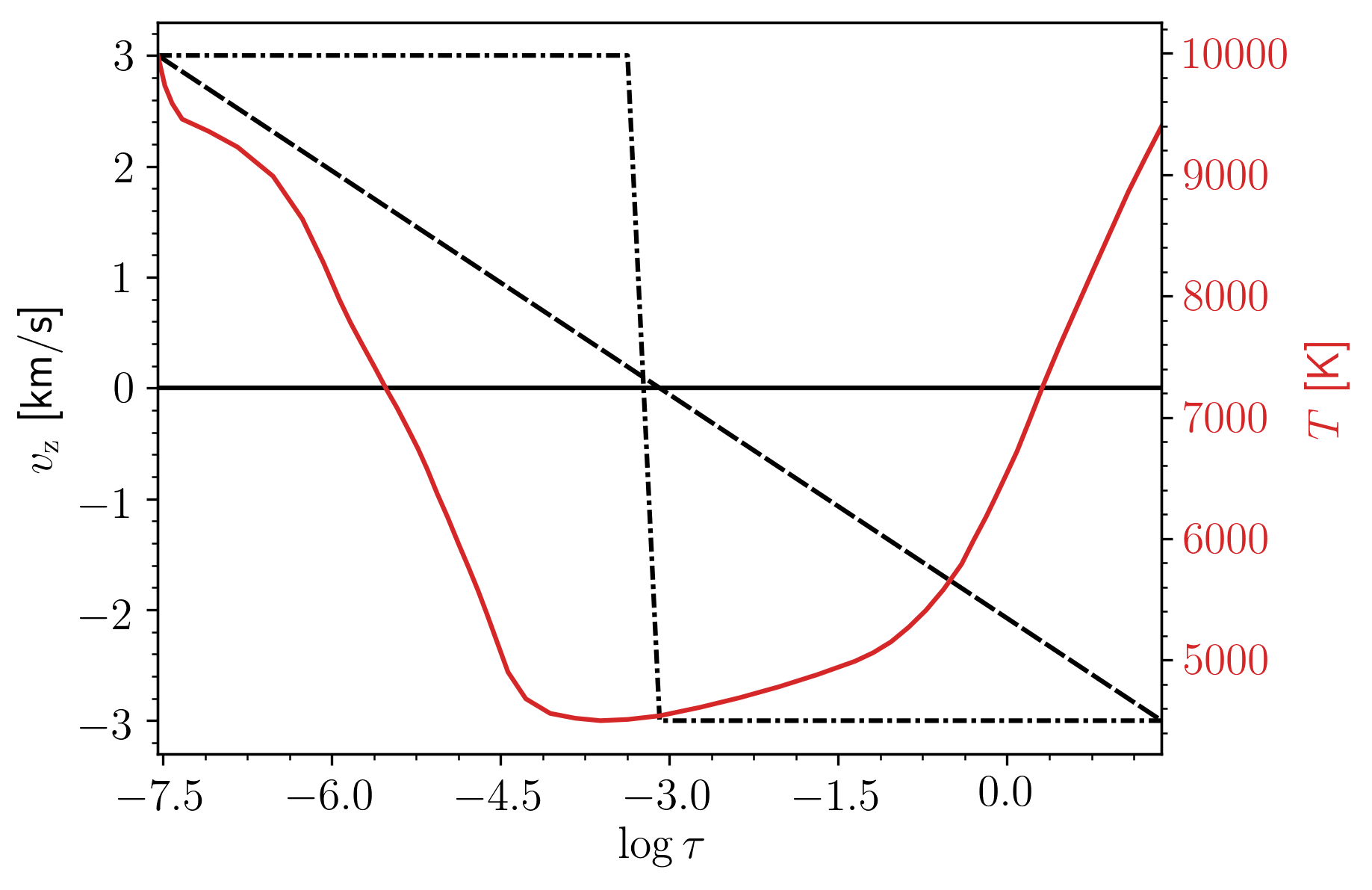}
    \caption{Two LOS velocity distributions added to the static FAL\,C atmospheric model are presented. The red curve illustrates the temperature stratification in the FAL\,C model.}
    \label{fig:vz}
\end{figure}

Since the depth-dependent plasma velocity can have a strong influence on the line profile affecting the reliability of the WFA, following \cite{rebeka} we considered two cases of vertical velocity distributions in the FAL\,C model with a constant magnetic field. The static atmosphere was used as the base model. First, we implemented a LOS velocity gradient with linear dependence on $\log \tau $, from -3 km/s at the bottom up to 3 km/s at the top. In the second scenario a velocity discontinuity is introduced at $\log \tau \approx -3$, i.e. at the height corresponding to temperature minimum where the core of the Mg\,I $b_2$ line is formed \citep{Quintero_mg}. The velocity distributions are given in Fig.\,\ref{fig:vz}.

The LOS magnetic field {was inferred from these atmospheric models with a Markov Chain Monte Carlo (MCMC) algorithm from the \texttt{emcee} Python package \citep{emcee}. Using the MCMC, which allows a proper assessment of uncertainties in the inferred parameters, we fit Eq.\,\ref{WFeq} to the synthetic Stokes $V$ spectra, one line at the time. To make this analysis more realistic, we convolved the synthesized spectra with a Gaussian corresponding to the spectral resolution $R=200\,000$ and added the wavelength-dependent noise $\sigma=10^{-3}$.

As the Fe\,I line is the blend of two lines, we used the index $s_\mathrm{V}$ describing line sensitivity to the circular polarization \citep{biblija} as the weighting coefficient. We assumed that both Fe\,I components are sensitive to the magnetic field at the same height. 

In Fig.\,\ref{fig:wf_validity} we compare the magnetic field of the model to the value inferred from each spectral line using the WFA for different scenarios of plasma flow. A disagreement appears when we use the WFA to infer the magnetic field from the atmosphere with a velocity discontinuity. The discontinuity affects the Fe\,I and Mg\,I line profiles, especially the shape of the line cores, and can lead to their splitting. This makes the results of the WFA unreliable, especially in the case of the Fe\,I line.

\begin{figure*}
    \includegraphics[width=\hsize]{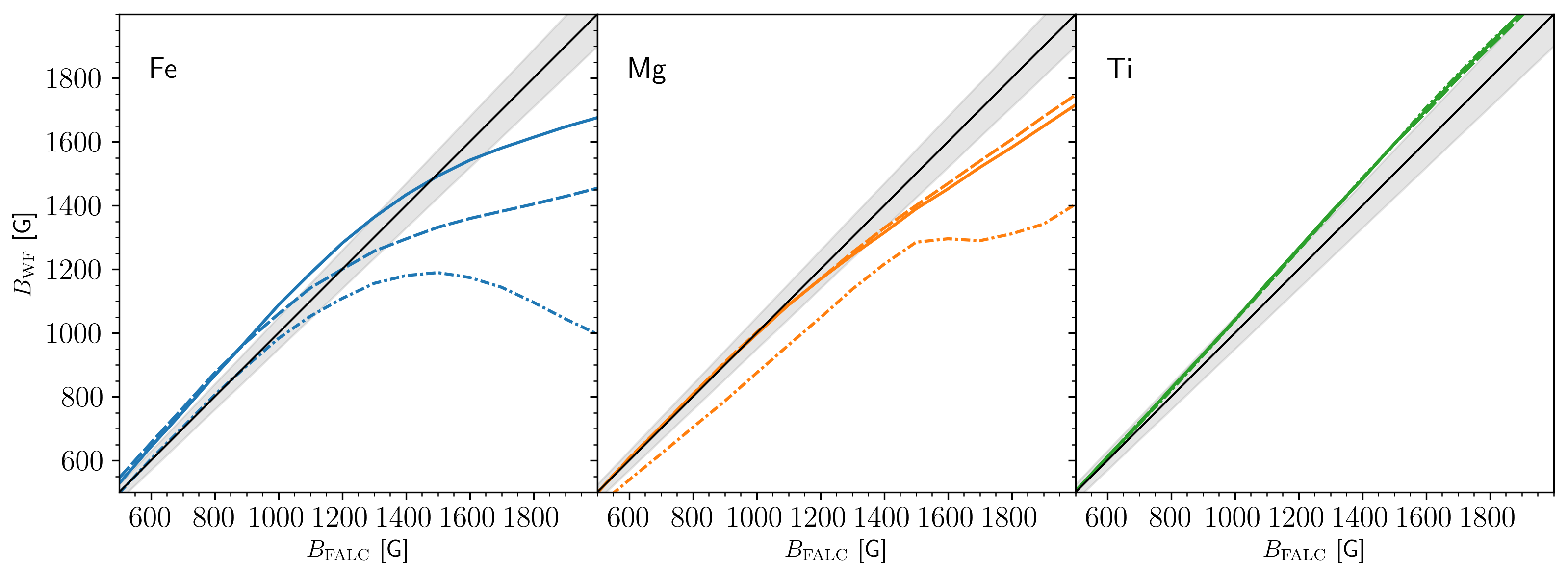}
    \caption{The comparison between the LOS magnetic field retrieved from the WFA and the LOS field strength of the model for each spectral line separately. Three regimes of vertical plasma flow are presented: static atmosphere (full line), smooth velocity gradient (dashed line) and the velocity discontinuity at $\log \tau \approx -3$ (dotted line). Shadowed region corresponds to the $5\%$ uncertainty.}
    \label{fig:wf_validity}
\end{figure*}

In the case of velocity gradient, the LOS magnetic fields are retrieved with $5\%$ accuracy for the field strengths up to around $1$\,kG in each line. For stronger fields the discrepancy is most prominent when the Fe\,I line is used (left panel of Fig.\,\ref{fig:wf_validity}), attributed to the difference in the formation heights of the two components. Because of the velocity gradient, each line of the blend is shifted by a different amount to shorter wavelengths, which leads to increasing differences between them. The Zeeman splitting produces further separation of the two Fe\,I lines. The resulting line profile of the blend can not be adequately described by the WFA, and there is a mismatch between the retrieved and the input magnetic field strengths. The difference in formation heights of Fe\,I components can also affect the inference of the LOS magnetic field from the atmosphere with a velocity discontinuity.

In the case of the Mg\,I line, we see that the presence of the velocity gradient does not significantly influence the magnetic field inference (middle panel of Fig.\,\ref{fig:wf_validity}). However, the discontinuity in the LOS velocity leads to significant differences between the WFA-inferred and the input field strengths.

The Ti\,I line shows a very good match between the WFA inferred and the model magnetic field up to $2$\,kG (right panel of Fig.\,\ref{fig:wf_validity}). This result is probably because of low Land\' e factor ($g_\mathrm{eff}=0.67$) that extends the applicability of the WFA to the strong magnetic fields. However, for the magnetic fields larger than $\sim1$\,kG, the retrieved field strength is slightly greater than the input one. We are not sure what is the cause of this disagreement. A similar effect was noticed for the Fe\,I line, but for weaker magnetic fields. The perfect match is achieved for the Ti\,I line for the magnetic field strengths less than $500$\,G.

The uncertainties in the inferred magnetic fields are $20-30$\,G for the continuum noise value $\sigma=10^{-3}$ and decrease to $2-3$\,G for $\sigma=10^{-4}$, while the best-fit values do not change with the noise. In our further analysis we consider only the wavelength-dependent noise with the continuum noise level of $\sigma=10^{-3}$.


In Fig.\,\ref{fig:bad_IV} we show Stokes $I$ and $V$ profiles calculated from the FAL\,C atmospheric model with the magnetic field of $1.6$\,kG for the three cases of the LOS velocity distribution. We chose to display the results for this value of magnetic field because the WFA failed to reproduce them entirely in all three cases. We also display the Stokes $V$ profiles obtained from  Eq.\,\ref{WFeq} with the best-fit $B_\mathrm{LOS}$. For the field strength of $1.6$\,kG the splitting in Fe\,I and Mg\,I lines is visible. With the added velocity distributions, the splitting becomes stronger and the mismatch between the weak-field profile and the synthetic one gets larger. Moreover, in the case of the velocity gradient there are asymmetries in the line profiles. Only the Ti\,I line does not show the Zeeman splitting or line asymmetry because of the velocity field.

\begin{figure*}
    \includegraphics[width=\hsize]{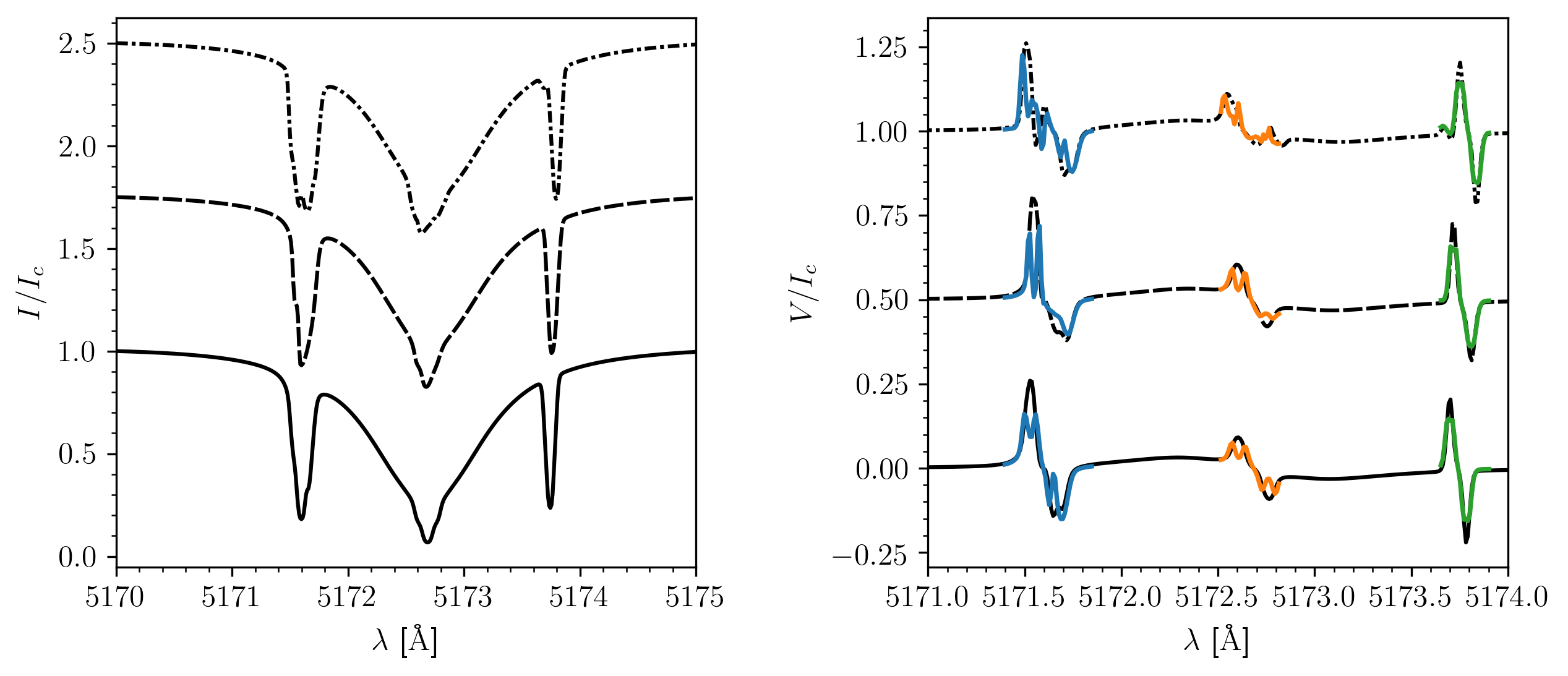}
    \caption{Stokes $I$ and $V$ line profiles from the FAL\,C model with the magnetic field strength of $1.6$\,kG. The profiles are given for different LOS velocity distributions: static atmosphere (full line), smooth velocity gradient (dashed), and the velocity discontinuity at $\log \tau \approx -3$ (dashed-dotted). Colored profiles represent the best fit of Stokes $V$ weak-field profiles for each line.}
    \label{fig:bad_IV}
\end{figure*}

In general, we have to assign the magnetic field value retrieved by the WFA to a specific depth. One way to do this is to use the response functions \citep[e.g.][]{Han06}. In the upper panels of Fig.\,\ref{fig:rf_bwf} we show the response functions of Stokes $V$ for the three considered lines to the magnetic field. These response functions are computed using the SNAPI code for FAL\,C atmosphere with uniform LOS magnetic field of $1$\,kG. We see that the Mg\,I line core is sensitive to the magnetic field in a large range of depths from the upper photosphere to the middle chromosphere. The Fe\,I and Ti\,I lines are sensitive mostly to the lower layers, spanning $\log \tau$ range from $-3$ to $-1$. 

Additionally, we calculated the response functions of the WFA magnetic field to the atmospheric magnetic field at different depths. These response functions (lower panel of Fig.\,\ref{fig:rf_bwf}) quantify how much the magnetic field at each depth in the atmosphere influences the WFA-inferred LOS magnetic field. The circular polarization in the Mg\,I line is formed over a broad range of heights, whereas the response functions for the Fe\,I and Ti\,I lines show a single peak suggesting that the LOS magnetic field inferred from the WFA can be related to a specific height. We estimated the Stokes $V$ formation height as the median value of these response functions. The calculated formation heights are $\log\tau_\mathrm{Fe}=-1.37$, $\log\tau_\mathrm{Mg}=-3.66$ and $\log\tau_\mathrm{Ti}=-1.35$. In the next section we will compare these formation heights with the values obtained from MURaM cubes that contain depth-dependent magnetic fields.



\begin{figure*}
    \includegraphics[width=\hsize]{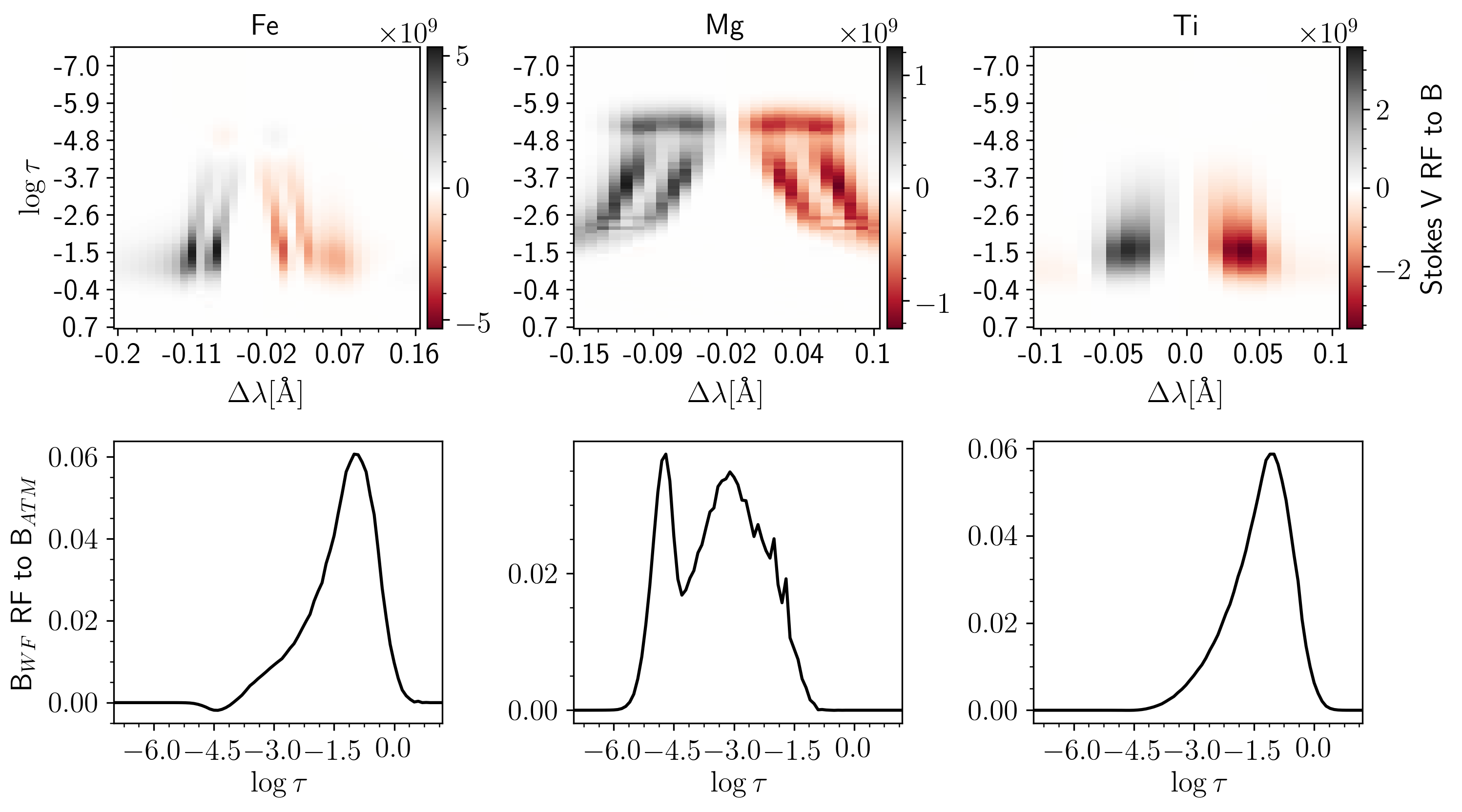}
    \caption{\textit{Upper panels:} Response functions of Stokes $V$ signal to the magnetic field in the FAL\,C model with the constant vertical $1$\,kG magnetic field. \textit{Lower panels:} Response function of weak-field estimation to the atmospheric magnetic field.}
    \label{fig:rf_bwf}
\end{figure*}

\section{Weak-field approximation validity in 3D atmospheres}
\label{sec:wfa_3d}

\subsection{Three dimensional atmosphere models}
\label{sec:atmospheres}

In general, solar magnetic fields are depth-dependent. To study the applicability of the WFA for inferring depth-dependent LOS magnetic fields from the Fe\,I, Mg\,I and Ti\,I lines around $5172$\,\AA\ we used two 3D models of the solar atmosphere resulting from two different MURaM simulation runs \citep{muram}. These are radiative-magneto-hydrodynamic simulations with the homogeneous magnetic field at the lower boundary equal to $50$\,G and $100$\,G, respectively. The models have horizontal extent of $6\times6$\,Mm and span $1.4$\,Mm in height, reaching roughly up to the temperature minimum / lower chromosphere.

We used the first MURaM cube with the field strength of $50$\,G at the lower boundary as the calibration cube from which we inferred Stokes $V$ formation heigths for the three lines (Sec. \ref{sec:calib}). Then, we used the second MURaM cube with the field strength $100$\,G at the lower boundary to test how good the LOS magnetic field is retrieved by the WFA. For this, we computed the spectra at the disk center ($\mu=1$) with a resolution of $10$\,m\AA. The synthetic profiles were convolved with the Gaussian corresponding to the spectral resolution $R=200\,000$ and we added the wavelength-dependent noise $\sigma=10^{-3}$. When calculating spectra from the MURaM 3D cube, we treated it as a series of 1D atmospheres, assuming that each vertical column of the cube is an independent atmosphere. Namely, we neglected the influence of the radiation coming from the neighboring atmospheres (so-called, 1.5D approach). Hereinafter, these individual atmospheres will be referred to as pixels.

In Fig.\,\ref{fig:fts_vs_synth} we compare the mean spectrum calculated from the first cube with that obtained from the FAL\,C model and the observed mean quiet Sun spectrum taken from \cite{bass2000}. The mean synthetic spectrum from the MURaM cube agrees well with the observed mean solar spectrum. There are small differences in the line core where the calculated intensity is slightly lower than in the observed spectrum. As the most of the line features are very well reproduced, we deem this agreement satisfactory for the upcoming discussion of the LOS magnetic field inference. 

\begin{figure}
    \includegraphics[width=\hsize]{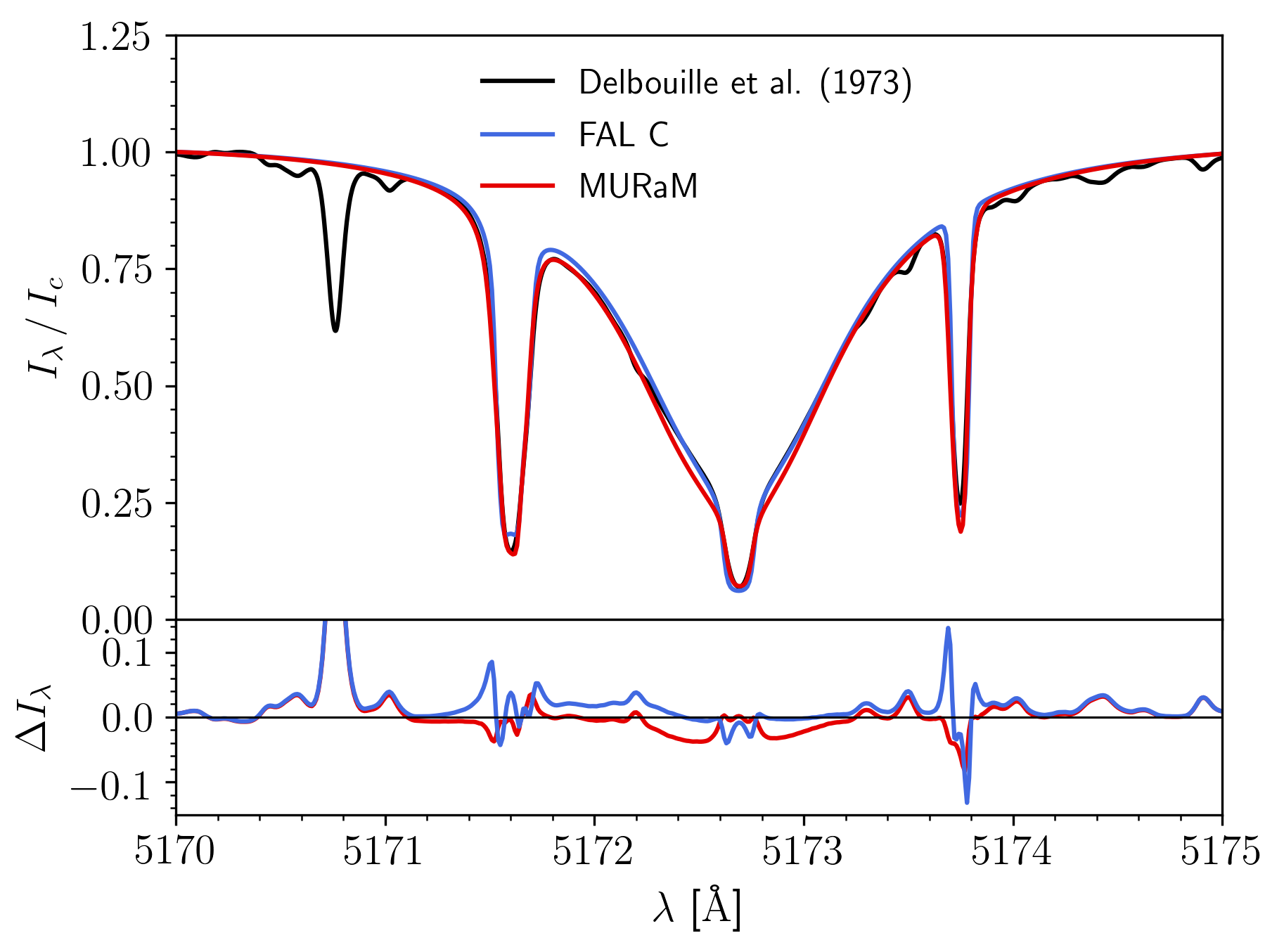}
    \caption{Comparison of the mean observed solar spectrum, the mean synthetic spectrum from a MURaM cube with the $50$\,G magnetic field at the lower boundary, and the synthetic spectrum from the FAL\,C model. At the lower subplot, we display the differences between the observed and the synthesized spectra.}
    \label{fig:fts_vs_synth}
\end{figure}

\subsection{Heights of formation}
\label{sec:calib}


When we infer magnetic field from the observed polarized spectrum using the WFA we need to relate the obtained value to a specific depth in the atmosphere, as the magnetic field changes with depth. Response functions give us an estimate of this formation depth, but for a specific atmospheric model. When analyzing the real data, the exact atmospheric structure for the observed pixel is unknown without performing a complete spectropolarimetric inversion. It is a complicated and computationally expensive process that might not always be feasible.

Instead, we aim to calibrate the WFA method using the MURaM cube with $50$\,G at the lower boundary. For that, we applied the WFA to the synthetic Stokes $V$ profiles of each of the three spectral lines and obtained three values of the LOS magnetic field for each pixel. Then, we compared these values to the known, depth-dependent, magnetic field at each pixel in the simulation cube. For each line we calculated the standard deviation of the difference between the inferred magnetic field and the model one at each depth point for all the considered pixels. The optical depth at which the standard deviation has its minimum value is taken to be the Stokes $V$ formation height. We use optical depth as we cannot infer the stratification of physical parameters on an absolute geometrical grid directly from the observations \citep[but see the recent paper by][]{Fritz}. We consider only the pixels with Stokes $V$ signal stronger than 0.2\% (normalized to the continuum Stokes $I$). These pixels mostly correspond to the intergranular lanes where the magnetic field is amplified by plasma motions and take up roughly $20\%$ of the total surface. The results presented here refer only to this subset of pixels.

\begin{figure}
    \includegraphics[width=\hsize]{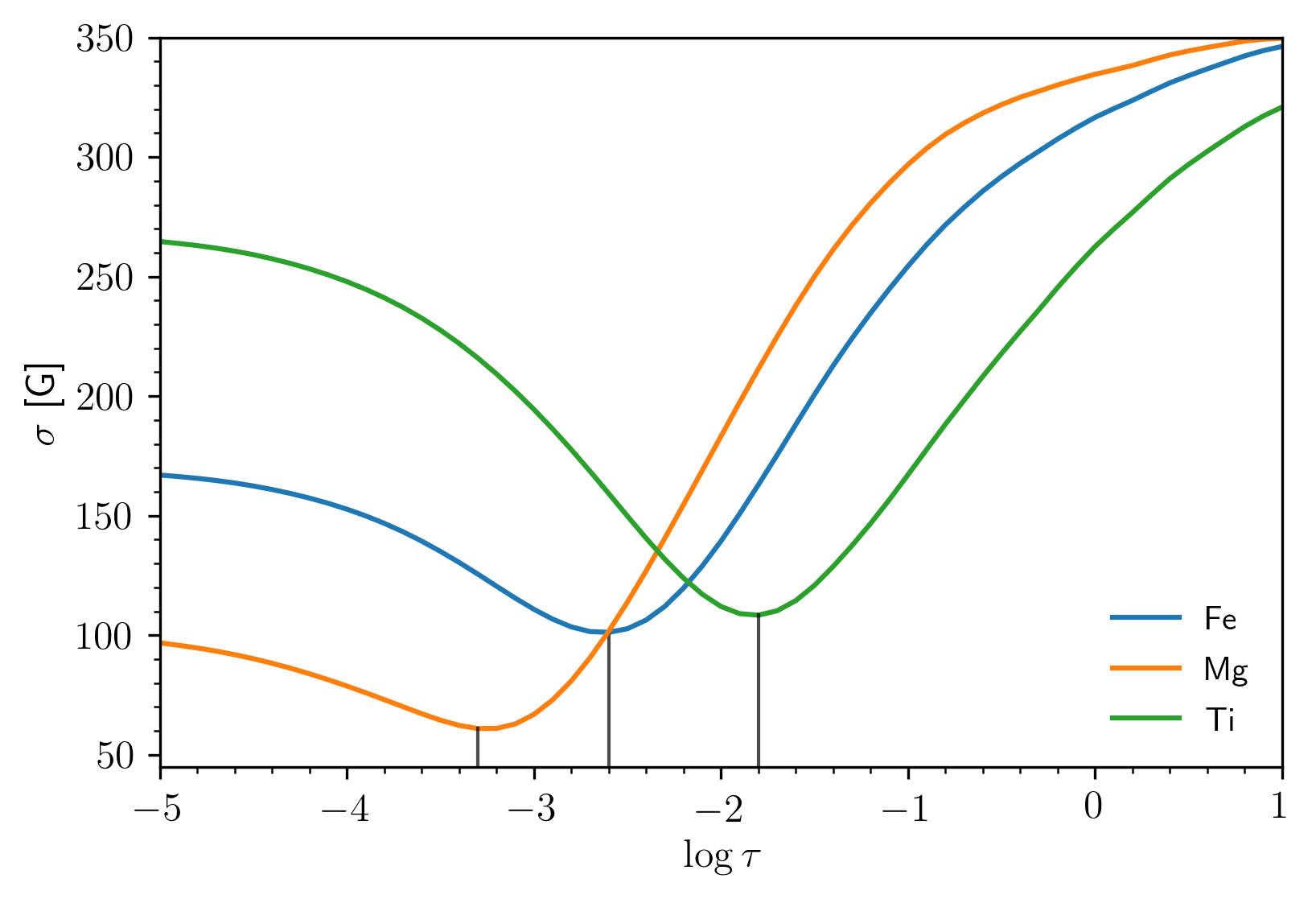}
    \caption{The distribution of standard deviation with optical depth from the MURaM cube with the $50$\,G magnetic field at the lower boundary. Vertical lines mark heights at which the minimum values of the standard deviation are reached.}
    \label{fig:std_pearson}
\end{figure}

Fig.\,\ref{fig:std_pearson} shows the distribution of the standard deviation with depth for each line. The minimum standard deviations for the three lines are at the following optical depths: $\log \tau_\mathrm{Fe} = -2.6$, $\log \tau_\mathrm{Mg} = -3.3$ and $\log \tau_\mathrm{Ti} = -1.8$. For the Mg\,I line this value is comparable to the one calculated from response function in the previous section using the FAL\,C model ($\log \tau=-3.66$). For the Ti line ($\log \tau=-1.35$ vs $-1.8$), and especially for the Fe line ($\log \tau=-1.37$ vs $-2.6$), these differences are larger.  The sources of this discrepancy can be different atmospheric models,  different estimation methods of the line formation heights, and in the case of the Fe\,I line the fact that it is a blend.


Fig.\,\ref{fig:mhd_slice_hofs} shows the Stokes $V$ formation height for each spectral line in a vertical slice of the MHD atmosphere with the $50$\,G magnetic field strength at the lower boundary. It can be seen that these three lines together provide information on the magnetic field strength over the height range of roughly 250\,km. The heights of formation estimated in this way should be taken as the effective Stokes $V$ formation heights for these spectral lines. The sensitivity to the magnetic field and the Stokes $V$ formation height for each of these lines are, however, expected to vary with atmospheric model.


\begin{figure*}
    \centering
    \includegraphics{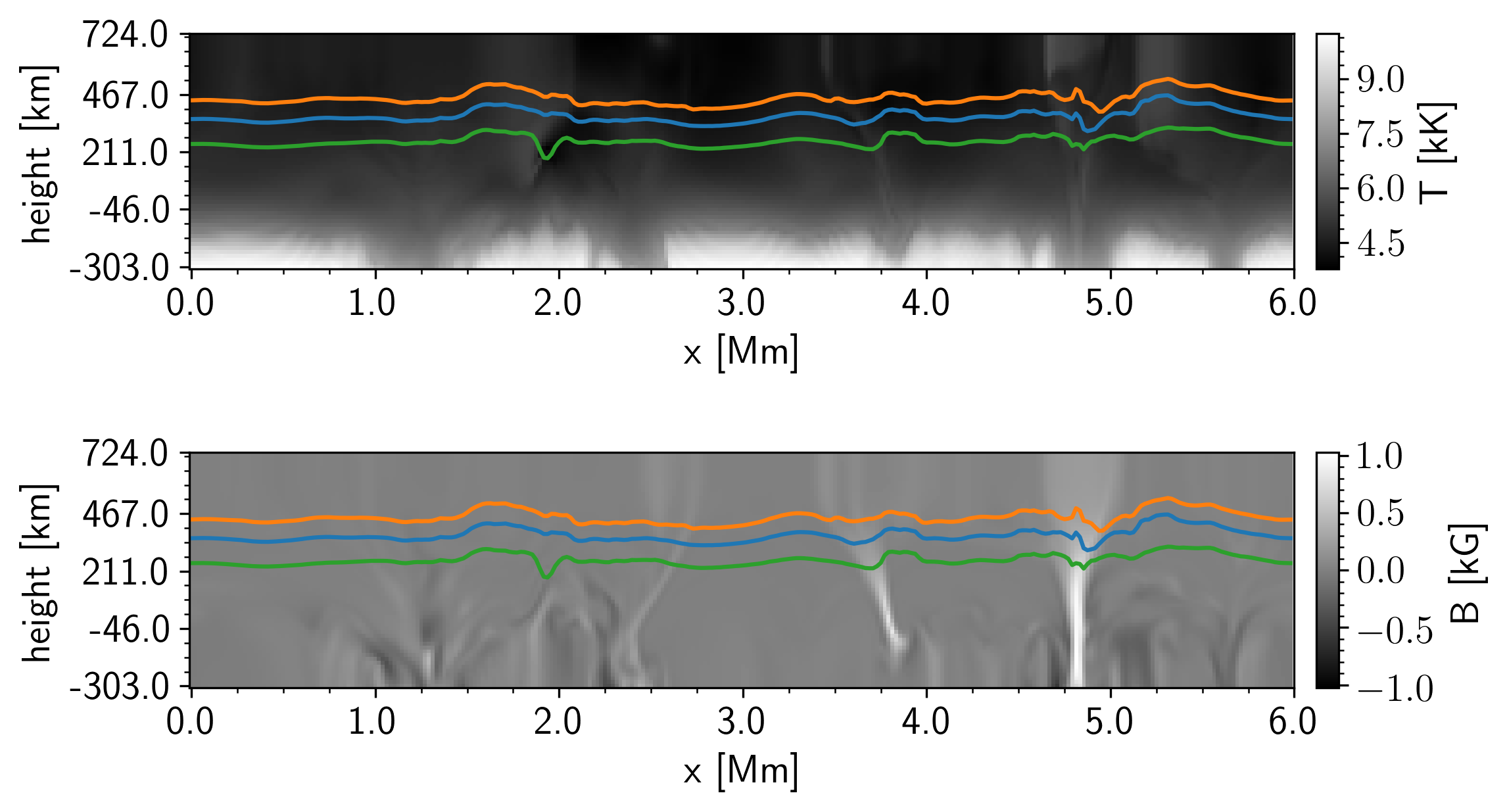}
    \caption{The Stokes $V$ formation height for spectral lines of Fe\,I (blue line), Mg\,I (orange line), and Ti\,I (green line) in the vertical slice from the MURaM cube with the $50$\,G magnetic field at the bottom boundary. The heights of formation are shown with respect to the temperature (upper panel) and to the LOS magnetic field strength (bottom panel).}
    \label{fig:mhd_slice_hofs}
\end{figure*}

 

We verified the inferred Stokes $V$ formation heights by applying the WFA to the polarized spectra calculated from the other MURaM cube. In Fig.\,\ref{fig:stokes} we show the emergent intensity, circular, and linear polarization from this MURaM cube, in the cores of Fe\,I, Mg\,I and Ti\,I lines. From the intensity map in the Fe\,I and Ti\,I lines, we see the reverse granulation effect in the higher photospheric layers. In the Mg\,I line, the granular structure disappears and we see even higher layers (temperature minimum and lower chromosphere). The Ti\,I line is formed in deeper layers than the Fe\,I line (evident also in Fig.\,\ref{fig:mhd_slice_hofs}). The Stokes $V$ structures are stronger and sharper when observed in the Fe\,I and Ti\,I lines, confirming that they sample lower layers than the Mg\,I line. The linear polarization signal is very low in all three lines. 

\begin{figure*}
    \resizebox{\hsize}{!}
    {\includegraphics{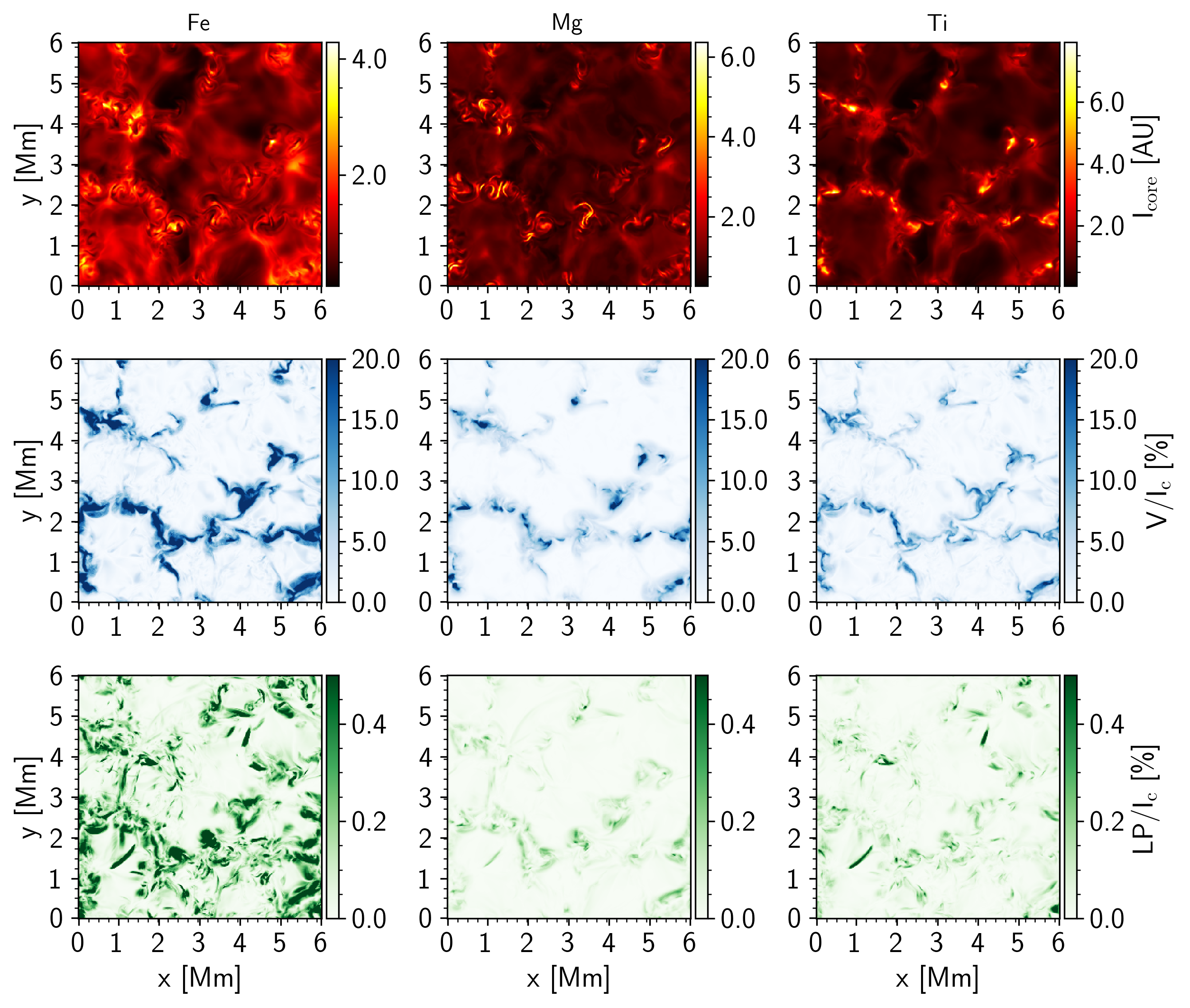}}
    \caption{The line core intensity (upper row), normalized to the spatially averaged value, the maximum circular polarization signal (middle row) and linear polarization signal (lower row) normalized to the mean quiet Sun spectrum, from the MURaM atmosphere with the $100$\,G magnetic field at the bottom boundary.}
    \label{fig:stokes}
\end{figure*}

In Fig.\,\ref{fig:2d_surface} we show the spatial distribution of the difference between the original magnetic field at the Stokes $V$ formation height and the magnetic field inferred by the WFA. In the case of Fe\,I and Ti\,I lines, the standard deviation of this difference is around $\sigma = 150$ G while for the Mg\,I line $\sigma = 40$ G. The largest discrepancies (red and orange points) are located at the central parts of the intergranular lanes. In these regions, the magnetic field can have large gradients. Also, the plasma flow direction is changing from horizontal to the vertical one. These effects limit the applicability of the weak-field approximation. 

In Fig.\,\ref{fig:scatter} we show scatter plots that compare the inferred magnetic field with the magnetic field from the MURaM cube at the Stokes $V$ formation height. These indicate that the WFA retrieves the vertical magnetic field very well for a wide range of magnetic field strengths.

\begin{figure*}
    \resizebox{\hsize}{!}
    {\includegraphics{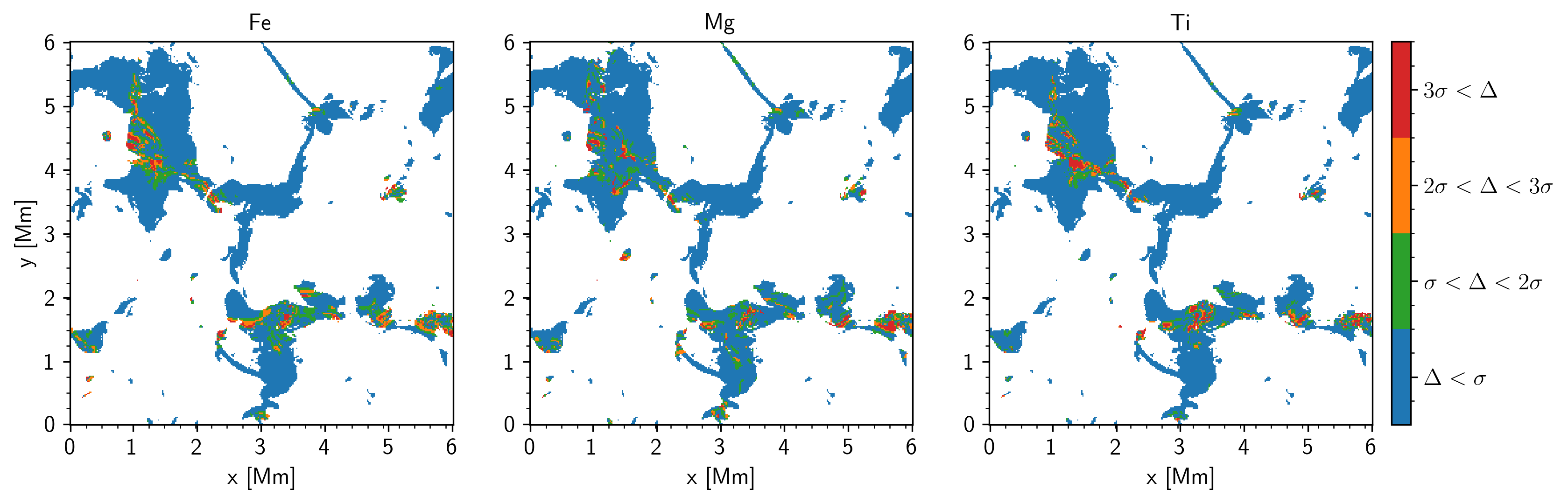}}
    \caption{The surface distribution of pixels that have Stokes $V$ signal stronger than $0.2\%$. We color-coded data on the plot depending on how dispersed are they from the mean value of the difference between the fields.}
    \label{fig:2d_surface}
\end{figure*}

\begin{figure*}
    \resizebox{\hsize}{!}
    {\includegraphics{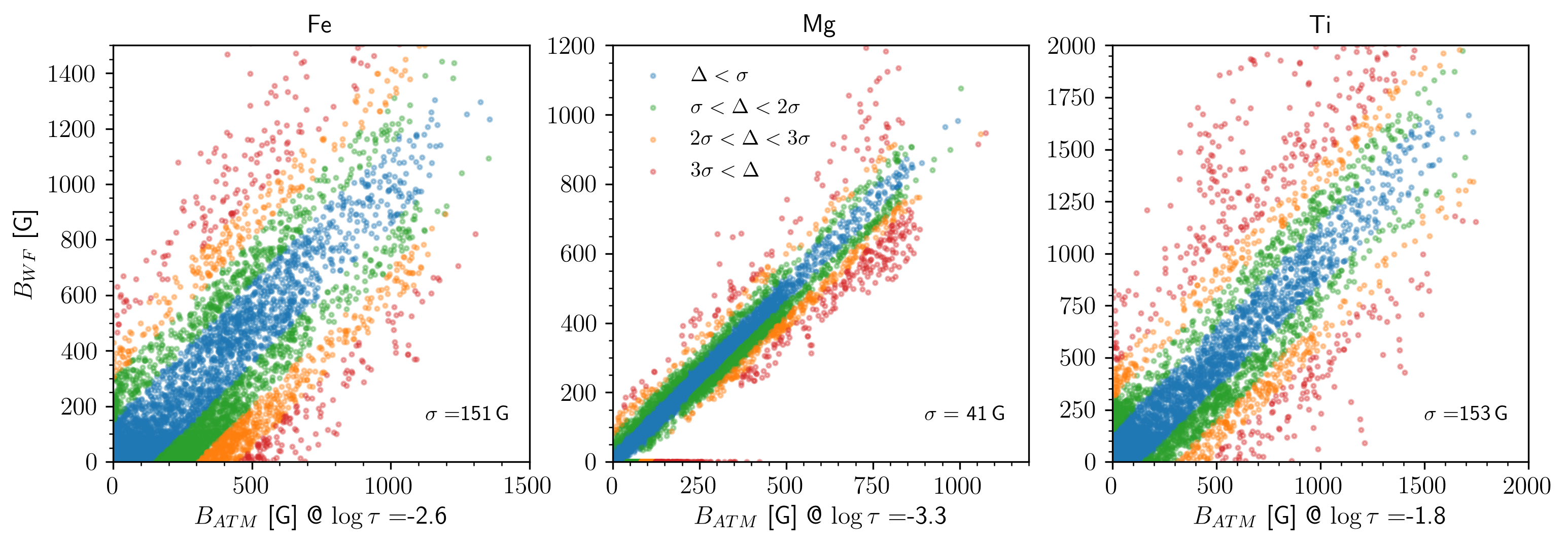}}
    \caption{Comparison of the weak-field and the MURaM cube LOS magnetic field at the height of formation for each line. Note the different ranges of magnetic field values on each panel. Each panel contains the standard deviation of the differences $B_\mathrm{WF}-B_\mathrm{ATM}$.  Results are from the MURaM cube with the $100$\,G magnetic field at the bottom boundary. Color-coding is the same as in Fig.\,\ref{fig:2d_surface}.}
    \label{fig:scatter}
\end{figure*}



\section{Conclusion}



We investigated and discussed how well we can retrieve a depth-dependent LOS magnetic field from the observations of the spectral region around the Mg\,I $b_2$ line, using the weak-field approximation (WFA). In addition to the Mg\,I line, we considered the Fe\,I (blend) and the Ti\,I photospheric lines sensitive to the magnetic field. Since the Mg\,I line is formed under NLTE conditions, we first constructed an atomic model with 12 levels, including continuum, based on the work of \cite{mauas}. Our model is a compromise between the simplicity necessary for fast synthesis and the ability to describe the observed line profiles well. We found that both the spectrum calculated from the FAL\,C atmospheric model and the mean spectrum from a quiet Sun MURaM simulation agree very well with the observations of the mean quiet Sun at the disk center. This is an interesting finding on its own because the MURaM simulations are typically used to model photospheric lines, which form lower than the Mg\,I line.

We tested the validity of the WFA in all three spectral lines in the semi-empirical FAL\,C atmospheric model. We included different distributions of the vertical velocity: static case, constant velocity gradient and velocity discontinuity. Applying the WFA to the Fe\,I line retrieves magnetic fields up to $1$\,kG within $5\%$ error for all three velocity distributions. For the Ti\,I line, the magnetic field is retrieved within the same error for the fields up to $2$\,kG. For the atmospheres with no velocity and constant velocity gradient, the Mg\,I line provides reliable ($<5\%$ error) magnetic field estimation for the magnetic fields up to $1.4$\,kG. However, when the velocity discontinuity is present, the WFA approximation yields significantly larger discrepancies.



Next, we considered the magnetic field inference from the synthetic line profiles calculated from a MURaM MHD atmosphere. We restricted the analysis to those pixels having the Stokes $V$ signal stronger than $0.2\%$ in all three lines. These pixels mostly correspond to the intergranular lanes, constituting approximately $20\%$ of the total number of pixels in the MURaM cube. First, we used a MHD atmosphere with the 50\,G magnetic field at the lower boundary to estimate the Stokes $V$ line formation height. We defined them as the optical depths where the standard deviation of the difference between the retrieved magnetic field and the original one is minimal. We found that the Stokes $V$ in the Mg\,I line is sensitive to regions around $\log \tau = -3.3$. For the Fe\,I and Ti\,I lines these depths are $\log \tau=-2.6$ and $\log \tau=-1.8$, respectively.

In order to test the estimated Stokes $V$ formation heights, we calculated the LOS magnetic field using the WFA from spectra calculated from another MURaM MHD atmosphere. We compared the inferred magnetic field with the original one at the corresponding Stokes $V$ formation heights and found a good agreement. In the pixels with steep vertical gradients in magnetic and/or velocity fields we find significant differences. 

We would also like to point out the work by \cite{regularized_wf} in which the WFA is applied to the SST observations of the Mg\,I $b_2$ line. The authors showed that even in the presence of a substantial noise ($\sigma \sim 10^{-2}$), the use of regularized WFA retrieved reliable values of the magnetic field. 

Note that these results should not be taken for granted. Namely, we did both the calibration and the test on the MHD atmospheres simulated with the same code (MURaM). Presumably, a weak-field estimate of the magnetic field will correspond to similar heights in both of these atmospheres. However, the good fit between the spectral line shapes calculated from the MURaM atmospheres and the solar observations encourages us to test this method on the real-life data. Our results can be used for fast estimation of the magnetic field strength or as initial values for more sophisticated spectropolarimetric inversions. This study does not include the estimation of the inclination and azimuth of the magnetic field, but the extension of this research with the approach similar to that of \cite{rebeka} is possible.


Upcoming facilities, such as the DKIST \citep{dkist} and the TuMag, will deliver so large quantities of data that classic spectropolarimetric inversions might not be computationally feasible. In that case, a quick estimation like the one presented in this paper can give us a valuable information of the depth dependence of the LOS magnetic field. In addition, these wavelengths are close to the maximum of the Planck curve at solar temperatures and also offer a good spatial resolution. Therefore, we advocate the use of this spectral region for studying the moderate magnetic fields in the photosphere and near the temperature minimum.

\begin{acknowledgements}
    Authors thank Rebecca Centeno for reading the draft version of the paper and providing valuable feedback and suggestions. We also thank the anonymous referee on the suggestions that made the paper much more clear and concise. 
    
    D.V. was funded by International Max Planck Research School (IMPRS) for Solar System Science at the University of Göttingen. This research is also funded by the Serbian Ministry of Science and Education, under the contract number 451-03-68/2022-14/200104. 
\end{acknowledgements}

\end{document}